\documentclass[graybox]{svmult}

\usepackage{amsmath,amssymb}
\usepackage{placeins}
\usepackage{hyperref} 
\usepackage{cite}
\usepackage{multirow}
\usepackage{subcaption}
\usepackage{url}

\usepackage{mathptmx} 
\usepackage{helvet} 
\usepackage{courier} 
\usepackage{graphicx} 
\usepackage{makeidx}
\usepackage{multicol} 
\usepackage{footmisc}

\begin{document}  

\title{Computing the Line Index of Balance Using Integer Programming Optimisation}

\titlerunning{Computing the Line Index of Balance}

\author{Samin Aref, Andrew J. Mason and Mark C. Wilson}

\institute{Samin Aref \at Department of Computer Science, University of Auckland\\ Auckland, Private Bag 92019, New Zealand \\ \email{sare618@aucklanduni.ac.nz}
	\and Andrew J. Mason \at Department of Engineering Science, University of Auckland
	\and Mark C. Wilson \at Department of Computer Science, University of Auckland}

\maketitle

\begin{abstract}\\
An important measure of signed graphs is the line index of balance which has applications in many fields. However, this graph-theoretic measure was underused for decades because of the inherent complexity in its computation which is closely related to solving NP-hard graph optimisation problems like MAXCUT.
We develop new quadratic and linear programming models to compute the line index of balance exactly. Using the Gurobi integer programming optimisation solver, we evaluate the line index of balance on real-world and synthetic datasets. The synthetic data involves Erd\H{o}s-R\'{e}nyi graphs, Barab\'{a}si-Albert graphs, and specially structured random graphs. We also use well known datasets from the sociology literature, such as signed graphs inferred from students' choice and rejection, as well as datasets from the biology literature including gene regulatory networks. 
The results show that exact values of the line index of balance in relatively large signed graphs can be efficiently computed using our suggested optimisation models. We find that most real-world social networks and some biological networks have small line index of balance which indicates that they are close to balanced.

\textbf{Keywords:} 
Integer programming,
Optimisation,
Frustration index,
Branch and bound,
Signed graphs,
Structural balance theory

\textbf{Mathematics Subject Classification (MSC 2010):}
05C22 15C22 90C09 90C11 90C90 90C35

\end{abstract}

\section{Introduction} \label{s:intro}

Graphs with positive and negative edges are referred to as \textit{signed graphs} \cite{zaslavsky_mathematical_2012} which are very useful in modelling the dual nature of interactions in various contexts. Graph-theoretic conditions \cite{harary_notion_1953,cartwright_structural_1956} of the structural balance theory \cite{heider_social_1944,harary_notion_1953} define the notion of \textit{balance} in signed graphs. If the vertex set of a signed graph can be partitioned into $k \leq 2$ subsets such that each negative edge joins vertices belonging to different subsets, then the signed graph is balanced \cite{cartwright_structural_1956}. For graphs that are not balanced, a distance from balance (a measure of partial balance \cite{aref2015measuring}) can be computed.

Among various measures is the \textit{frustration index} that indicates the minimum number of edges whose removal results in balance \cite{abelson_symbolic_1958,harary_measurement_1959,zaslavsky_balanced_1987}. This number was originally proposed in oblique form and referred to as \textit{complexity} by Abelson et al.\ \cite{abelson_symbolic_1958}. One year later, Harary proposed the same idea much more clearly with the name \textit{line index of balance} \cite{harary_measurement_1959}. More than two decades later, Toulouse used the term \textit{frustration} to discuss the minimum energy of an Ising spin glass model \cite{toulouse_theory_1987}. Zaslavsky has made a connection between the line index of balance and spin glass concepts and introduced the name frustration index \cite{zaslavsky_balanced_1987}. We use both names, line index of balance and frustration index, interchangeably in this chapter.

\section{Literature review}\label{s:literature}

Except for a normalised version of the frustration index \cite{aref2015measuring}, measures of balance used in the literature \cite{cartwright_structural_1956,norman_derivation_1972,terzi_spectral_2011,kunegis_applications_2014,estrada_walk-based_2014} do not satisfy key axiomatic properties \cite{aref2015measuring}. Using cycles \cite{cartwright_structural_1956,norman_derivation_1972}, triangles \cite{terzi_spectral_2011,kunegis_applications_2014}, Laplacian matrix eigenvalues \cite{kunegis_spectral_2010}, and closed-walks \cite{estrada_walk-based_2014} to evaluate distance from balance has led to conflicting observations \cite{leskovec_signed_2010, facchetti_computing_2011, estrada_walk-based_2014}. 

Besides applications as a measure of balance, the frustration index is a key to frequently stated problems in several fields of research \cite{aref2017balance}. In biology, optimal decomposition of biological networks into monotone subsystems is made possible by calculating the line index of balance \cite{iacono_determining_2010}. In finance, portfolios whose underlying signed graph has negative edges and a frustration index of zero have a relatively low risk \cite{harary_signed_2002}. In physics, the line index of balance provides the minimum energy state of atomic magnets \cite{kasteleyn_dimer_1963,Sherrington, Barahona1982}. In international relations, alliance and antagonism between countries can be analysed using the line index of balance \cite{patrick_doreian_structural_2015}. In chemistry, bipartite edge frustration indicates the stability of fullerene, a carbon allotrope \cite{doslic_computing_2007}. For a discussion on applications of the frustration index, one may refer to \cite{aref2017balance}.

Detecting whether a graph is balanced can be solved in polynomial time \cite{hansen_labelling_1978,harary_simple_1980,zaslavsky1983signed}. However, calculating the line index of balance in general graphs is an NP-hard problem equivalent to the ground state calculation of an unstructured Ising model \cite{mezard2001bethe}. Computation of the line index of balance can be reduced from the graph maximum cut (MAXCUT) problem, in the case of all negative edges, which is known to be NP-hard \cite{huffner_separator-based_2010}.

Similar to MAXCUT for planar graphs \cite{hadlock}, the line index of balance can be computed in polynomial time for planar graphs \cite{katai1978studies}. Other special cases of related problems can be found among the works of Hartmann and collaborators who have suggested efficient algorithms for computing ground state in 3-dimensional spin glass models \cite{hartmann2015matrix} improving their previous contributions in 1-, 2-, and 3-dimensional \cite{hartmann2014exact,hartmann2011ground,hartmann2013information} spin glass models. Recently, they have used a method for solving 0/1 optimisation models to compute the ground state of 3-dimensional models containing up to $268^3$ nodes \cite{hartmann2016revisiting}. 

A review of the literature shows 5 algorithms suggested for computing the line index of balance between 1963 and 2002. The first algorithm \cite[pages 98-107]{flament1963applications} is developed specifically for complete graphs. It is a naive algorithm that requires explicit enumeration of all possible combinations of sign changes that may or may not lead to balance. With a run time exponential in the number of edges, this is clearly not practical for graphs with more than 8 nodes that require billions of cases to be checked. The second algorithm is an optimisation method suggested by \cite{hammer1977pseudo}. This method is based on solving an unconstrained binary quadratic model. We will discuss a model of this type in Subsection \ref{ss:ubqp} and other more efficient models later in this chapter. The third computation method is an iterative algorithm suggested in \cite[algorithm 3, page 217]{hansen_labelling_1978}. The iterative algorithm is based on removing edges to eliminate negative cycles of the graph and only provides an upper bound on the line index of balance. A fourth method suggested by Harary and Kabell \cite[page 136]{harary_simple_1980} is based on extending a balance detection algorithm. This method is inefficient according to Bramsen \cite{bramsen2002further} who in turn suggests an iterative algorithm with a run time that is exponential in the number of nodes. Using Bramsen's suggested method for a graph with 40 nodes requires checking trillions of cases to compute the line index of balance which is clearly impractical. Doreian and Mrvar have recently attempted computing the line index of balance using a polynomial time algorithm \cite{patrick_doreian_structural_2015}. However, our computations on their data show that their solutions are not optimal and thus do not give the line index of balance.

This review of literature shows that computing the line index of balance in general graphs lacks extensive and systematic investigation.

\subsection*{Our contribution} \label{ss:contrib}
We provide an efficient method for computing the line index of balance in general graphs of the sizes found in many application areas. Starting with a quadratic programming model based on signed graph switching equivalents, we suggest several optimisation models. We use powerful mathematical programming solvers like Gurobi to solve the optimisation models. 

This chapter begins with the preliminaries in Section \ref{s:prelim}. Three mathematical programming models are developed in Section \ref{s:qmodel}. The results on synthetic data are provided in Section \ref{s:results}. Numerical results on real social and biological networks are provided in Section \ref{s:real} including graphs with up to 3215 edges. Section \ref{s:conclu} summarises the key highlights of the research. 

\section{Preliminaries} \label{s:prelim}

\subsection{Basic notation} \label{s:problem}
We consider undirected signed networks $G = (V,E,\sigma)$. The set of nodes is denoted by $V$, with $|V| = n$. $E$ is the set of edges that is partitioned into the set of positive edges $E^+$ and the set of negative edges $E^-$ with $|E^-|=m^-$, $|E^+|=m^+$, and $|E|=m=m^- + m^+$. The sign function, denoted by $\sigma$, is a mapping of edges to signs $\sigma: E\rightarrow\{-1,+1\}^m$. We represent the $m$ undirected edges in $G$ as ordered pairs of vertices $E = \{e_1, e_2, ..., e_m\} \subseteq \{ (i,j) \mid i,j \in V , i<j \}$, where a single edge $e_k$ between nodes $i$ and $j$, $i<j$, is denoted by $e_k=(i,j) , i<j$. We denote the graph density by $\rho= 2m/(n(n-1))$. The entries of the symmetric adjacency matrix $\textbf{A}=(a_{ij})$ are defined in \eqref{eq1}. 
\begin{equation}\label{eq1}
a{_i}{_j} =
\left\{
\begin{array}{ll}
\sigma_{(i,j)} & \mbox{if } (i,j) \in E \ \text{or} \ (j,i) \in E \\
0 & \mbox{if } (i,j)\notin E
\end{array}
\right.
\end{equation}

The number of positive (negative) edges connected to the node $i \in V$ is the positive (negative) degree of the node and is denoted by $d^+ {(i)}$ ($d^- {(i)}$). The net degree of a node is defined by $d^+ {(i)} -d^- {(i)}$.

A \emph{walk} of length $k$ in $G$ is a sequence of nodes $v_0,v_1,...,v_{k-1},v_k$ such that for each $i=1,2,...,k$ there is an edge between $v_{i-1}$ and $v_i$. If $v_0=v_k$, the sequence is a \emph{closed walk} of length $k$. If the nodes in a closed walk are distinct except for the endpoints, it is a \emph{cycle} of length $k$. The \emph{sign} of a cycle is the product of the signs of its edges. A balanced graph is one with no negative cycles \cite{cartwright_structural_1956}.

\subsection{Node colouring and frustration count}

For each signed graph $G=(V, E, \sigma)$, we can partition $V$ into two sets, denoted $X \subseteq V$ and $\bar X=V \backslash X$. We think of $X$ as specifying a colouring of the nodes, where each node $i \in X$ is coloured black, and each node $i \in \bar X$ is coloured white. 

We let $x_i$ denote the colour of node $i \in V$ under $X$, where $x_i=1$ if $i \in X$ and $x_i=0$ otherwise. We say that an edge $(i,j) \in E$ is {\em frustrated} under $X$ if either edge $(i,j)$ is a positive edge (\ $(i,j) \in E^+$) but nodes $i$ and $j$ have different colours ($x_i \ne x_j$), or edge $(i,j)$ is a negative edge (\ $(i,j) \in E^-$) but nodes $i$ and $j$ share the same colour ($x_i = x_j$). We define the {\em frustration count} $f_G(X)$ as the number of frustrated edges under $X$: $$f_G(X) = \sum_{(i,j) \in E} f_{ij}(X)$$
where for $(i,j) \in E$:
\begin{equation} \label{eq1.1}
f_{ij}(X)=
\begin{cases}
0, & \text{if}\ x_i = x_j \text{ and } (i,j) \in E^+ \\
1, & \text{if}\ x_i = x_j \text{ and } (i,j) \in E^- \\
0, & \text{if}\ x_i \ne x_j \text{ and } (i,j) \in E^- \\
1, & \text{if}\ x_i \ne x_j \text{ and } (i,j) \in E^+. \\
\end{cases}
\end{equation}

The frustration index $L(G)$ of a graph $G$ can be found by finding a subset $X^* \subseteq V$ of $G$ that minimises the frustration count $f_G(X)$, i.e., solving Eq.\ \eqref{eq1.2}.

\begin{equation} \label{eq1.2}
L(G)=\min_{X \subseteq V}f_G(X)\
\end{equation}

\subsection{Minimum deletion set and switching function}\label{ss:state}

For each signed graph, there are sets of edges, called \textit{deletion sets}, whose deletion results in a balanced graph. A minimum deletion set $E^*\subseteq E$ is a deletion set with the minimum size. The frustration index $L(G)$ equals the size of a minimum deletion set: $L(G)=|E^*|$. 

We define the \emph{switching function} $g$ operating over a set of vertices, called the \textit{switching set}, $X\subseteq V$ as follows in \eqref{eq2}.
\begin{equation} \label{eq2}
\sigma^g _{(i,j)}=
\left\{
\begin{array}{rl}
\sigma_{(i,j)} & \mbox{if } {i,j}\in X \ \text{or} \ {i,j}\notin X \\
-\sigma_{(i,j)} & \mbox{if } (i \in X \ \text{and} \ j\notin X) \ \text{or} \ (i \notin X \ \text{and} \ j \in X)
\end{array}
\right.
\end{equation}
The graph resulting from applying switching function $g$ to signed graph $G$ is called $G$'s \textit{switching equivalent} and denoted by $G^g$. The switching equivalents of a graph have the same value of the frustration index, i.e. $L{(G^g)}=L(G) \forall g$ \cite{zaslavsky_matrices_2010}. It is straightforward to prove that the frustration index is equal to the minimum number of negative edges in $G^g$ over all switching functions $g$. An immediate result is that any balanced graph can switch to an equivalent graph where all the edges are positive \cite{zaslavsky_matrices_2010}. Moreover, in a switched graph with the minimum number of negative edges, called a \textit{negative minimal graph} and denoted by $G^{g^*}$, all vertices have a non-negative net degree. In other words, if $m^- {(G^g)}=L(G)$ then every vertex $ i^g $ in switched graph $G^g$ satisfies $d^- {(i^g)} \leq d^+ {(i^g)}$. 

\subsection{Upper bounds for the line index of balance}\label{ss:bounds}
An obvious upper bound for the line index of balance is $L(G)\leq m^-$ which states the result that removing all negative edges gives a balanced graph. Recalling that acyclic signed graphs are balanced, the circuit rank of the graph can also be considered as an upper bound for the frustration index \cite[p. 8]{flament1970equilibre}. Circuit rank, also known as the cyclomatic number, is the minimum number of edges whose removal results in an acyclic graph.

Petersdorf \cite{petersdorf_einige_1966} proves that among all sign functions for complete graphs with $n$ nodes, assigning negative signs to all the edges, i.e. putting $\sigma: E\rightarrow\{-1\}^m$, gives the maximum value of the frustration index which equals $\lfloor (n-1)^2/4 \rfloor$. 
Petersdorf's proof confirms a conjecture by Abelson and Rosenberg\cite{abelson_symbolic_1958} that is also proved in \cite{tomescu_note_1973} and further discussed in \cite{akiyama_balancing_1981}.

Akiyama et al.\ provide results indicating that the frustration index of signed graphs with $n$ nodes and $m$ edges is bounded by $m/2$ \cite{akiyama_balancing_1981}. They also show that the frustration index of signed graphs with $n$ nodes is maximum in all complete graphs with no positive 3-cycles and is bounded by $\lfloor (n-1)^2/4 \rfloor$ \cite[Theorem 1]{akiyama_balancing_1981}. 
This group of graphs also contains complete graphs with nodes that can be partitioned into two classes such that all positive edges connect nodes from different classes and all negative edges connect nodes belonging to the same class \cite{tomescu_note_1973}. 
Akiyama et al.\ refer to these graphs as \textit{antibalanced} \cite{akiyama_balancing_1981} which is a term coined by Harary in \cite{harary_structural_1957} and also discussed in \cite{zaslavsky_matrices_2010}. 

\section{Mathematical programming models} \label{s:qmodel}

In this section, we formulate three mathematical programming models in \eqref{eq4}, \eqref{eq7}, and \eqref{eq8} to calculate the frustration index by optimizing an objective function formed using integer variables.

\subsection{A quadratically constrained quadratic programming model}

We formulate a mathematical programming model in Eq.\ \eqref{eq4} to maximise $Z_1$, the sum of entries of $\textbf{A}^g$, the adjacency matrix of the graph switched by $g$, over different switching functions. Bearing in mind that the frustration index is the number of negative edges in a negative minimal graph, $L(G)=m^{-}{(G^{g^*})}$, then maximising $Z_1$ will effectively calculate the line index of balance. We use decision variables, $y_{i} \in \{-1,1\}$ to define node colours. Then $X=\{i \mid y_{i}=1\}$ gives the black-coloured nodes (alternatively nodes in the switching set). The restriction $y_{i} \in \{-1,1\}$ for the variables is formulated by $n$ quadratic constraints $y^2_{i}=1$. Note that the switching set $X=\{i \mid y_{i}=1\}$ creates a negative minimal graph with the adjacency matrix entries given by $a_{ij} y_{i} y_{j}$. The model can be represented as Eq.\ \eqref{eq4} in the form of a continuous quadratically constrained quadratic programming (QCQP) model with $n$ decision variables and $n$ constraints.

\begin{equation}\label{eq4}
\begin{split}
\max_{y_i} Z_1 &= \sum\limits_{i \in V} \sum\limits_{j \in V} a_{ij} y_{i} y_{j} \\
\text{s.t.} \quad
y^2_{i}&=1 \quad \forall i \in V
\end{split}
\end{equation}

Maximising $\sum_{i \in V} \sum _{j \in V} a_{ij} y_{i} y_{j}$ is equivalent to minimising $m^{-}{(G^{g^*})}=|\{(i,j) \in E: a_{ij} y_{i} y_{j}=-1\}|$. The optimal value of the objective function, $Z_1^*$, is equal to the sum of entries in the adjacency matrix of a negative minimal graph which can be represented by $Z_1^*= 2m^{+}{(G^{g^*})} - 2m^{-}{(G^{g^*})} = 2m - 4L(G)$. Therefore, the graph frustration index can be calculated by $L(G)= (2m - Z_1^*)/4$.

While the model expressed in \eqref{eq4} is quite similar to the non-linear energy function minimization model used in \cite{facchetti_computing_2011, facchetti2012exploring ,esmailian_mesoscopic_2014, ma_memetic_2015} and the Hamiltonian of Ising models with $\pm 1$ interactions \cite{Sherrington}, the feasible region in model \eqref{eq4} is neither convex nor a second order cone. Therefore, the QCQP model in \eqref{eq4} only serves as an easy-to-understand optimisation model clarifying the node colouring (alternatively selecting nodes to switch) and how it relates to the line index of balance.

\subsection{An unconstrained binary quadratic programming model}\label{ss:ubqp}

The optimisation model \eqref{eq4} can be converted into an unconstrained binary quadratic programming (UBQP) model \eqref{eq7} by changing the decision variables into binary variables $y_{i}=2x_{i}-1$ where $x_{i} \in \{0,1\}$. Note that the binary variables, $x_{i}$, define the black-coloured nodes $X=\{i \mid x_{i}=1\}$ (alternatively, nodes in the switching set). The optimal solution represents a subset $X^* \subseteq V$ of $G$ that minimises the resulting frustration count. 

Furthermore, by substituting $y_{i}=2x_{i}-1$ into the objective function in \eqref{eq4} we get \eqref{eq5}. The terms in the objective function can be modified as shown in \eqref{eq5}--\eqref{eq6} in order to have an objective function whose optimal value, $Z^*_2$, equals $L(G)$.
\begin{equation}\label{eq5}
\begin{split}
Z_1 &= \sum\limits_{i \in V} \sum\limits_{j \in V} (4a_{ij} x_{i} x_{j} - 2 x_{i} a_{ij} - 2 x_{j} a_{ij} + a_{ij}) \\
 &= \sum\limits_{i \in V} \sum\limits_{j \in V} (4 a_{ij} x_{i} x_{j} - 4 x_{i} a_{ij}) + (2m - 4m^{-}_{(G)})
\end{split}
\end{equation}

\begin{equation}\label{eq6}
Z_2 = (2m - Z_1)/4
\end{equation}

Note that the binary quadratic model in Eq.\ \eqref{eq7} has $n$ decision variables and no constraints.

\begin{equation}\label{eq7}
\begin{split}
\min_{x_i} Z_2 &= \sum\limits_{i \in V} \sum\limits_{j \in V} (a_{ij}x_{i} - a_{ij}x_{i}x_{j})  + m^{-}_{(G)}\\
\text{s.t.} \quad x_{i} &\in \{0,1\} \ i \in V
\end{split}
\end{equation}

The optimal value of the objective function in Eq.\ \eqref{eq7} represents the frustration index directly as shown in \eqref{eq7.5}.

\begin{equation}\label{eq7.5}
Z_2^* = (2m - Z^*_1)/4 = (2m - (2m - 4L(G)))/4 = L(G)
\end{equation}

The objective function in Eq.\ \eqref{eq7} can be interpreted as initially starting with $m^{-}_{(G)}$ and then adding 1 for each positive frustrated edge (positive edge with different endpoint colours) and -1 for each negative edge that is not frustrated (negative edge with different endpoint colours). This adds up to the total number of frustrated edges.

\subsection{The linear programming model}

The linearised version of \eqref{eq7} is formulated in Eq.\ \eqref{eq8}. The objective function of \eqref{eq7} is first modified as shown in Eq.\ \eqref{eq7.6} and then its non-linear term $x_{i} x_{j}$ is replaced by $|E|$ additional binary variables $x_{ij}$. The new decision variables $x_{ij}$ are defined for each edge $(i,j) \in E$ and take value 1 whenever $x_{i}=x_{j}=1$ and $0$ otherwise.
Note that $d_{i}=\sum_{j \in V}a_{ij}$ is a constant that equals the net degree of node $i$.

\begin{equation}\label{eq7.6}
\begin{split}
Z_2 &= \sum\limits_{i \in V} \sum\limits_{j \in V} a_{ij}x_{i} - \sum\limits_{i \in V} \sum\limits_{j \in V}a_{ij}x_{i}x_{j}  + m^{-}_{(G)}\\
    &= \sum\limits_{i \in V} x_{i} \sum\limits_{j \in V}a_{ij} - \sum\limits_{i \in V} \sum\limits_{j \in V, j>i} 2a_{ij}x_{i}x_{j}  + m^{-}_{(G)}\\
    &=\sum\limits_{i \in V} x_{i} d_{i} - \sum\limits_{i \in V} \sum\limits_{j \in V, j>i} 2a_{ij}x_{i}x_{j}  + m^{-}_{(G)}
\end{split}
\end{equation}

The dependencies between the $x_{ij}$ and $x_{i},x_{j}$ values are taken into account by considering a constraint for each new variable. Therefore, the 0/1 linear model has $n+m$ variables and $m$ constraints as it follows in \eqref{eq8}.

\begin{equation}\label{eq8}
\begin{split}
\min_{x_i, x_{ij}} Z_2 &= \sum\limits_{i \in V} d_{i}x_{i}  - \sum\limits_{(i,j) \in E} 2a_{ij}x_{ij}  + m^{-}_{(G)}\\
\text{s.t.} \quad
x_{ij} &\leq (x_{i}+x_{j})/2 \quad \forall (i,j) \in E^+ \\
x_{ij} &\geq x_{i}+x_{j}-1 \quad \forall (i,j) \in E^- \\
x_{i} &\in \{0,1\} \ i \in V\\
x_{ij} &\in \{0,1\} \ (i,j) \in E
\end{split}
\end{equation}

\subsection{Additional constraints for the linear programming model}

The structural properties of the model allow us to restrict the model by adding additional valid inequalities. Valid inequalities are utilised by our solver Gurobi as additional non-core constraints that are kept aside from the core constraints of the model. Upon violation by a solution, valid inequalities are efficiently pulled in to the model. Pulled-in valid inequalities cut away a part of the feasible space and restrict the model. Additional restrictions imposed on the model can often speed up the solver algorithm if they are valid and useful \cite{Klotzpractical}. Properties of the optimal solution can be used to determine these additional constraints. Properties observed in negative minimal graphs include the nonnegativity of net degree of the nodes and negation states of the edges making a cycle.

An obvious structural property of the nodes in a negative minimal graph, $G^{g^*}$, is that their net degrees are always non-negative, i.e., $d^+ {(i^{g^*})} - d^- {(i^{g^*})} \geq 0 \quad \forall i \in V $. Equivalently, a node $i$ should be given a colour that minimises the number of frustrated edges connected to it. This can be proved by contradiction using the definition of the switching function \eqref{eq2}. To see this, assume a node in a negative minimal graph has a negative degree. It follows that the negative edges connected to the node outnumber the positive edges. Therefore, switching the node decreases the total number of negative edges in a negative minimal graph which is a contradiction.

This structural property can be formulated as constraints in the problem. A \textit{net-degree constraint} can be added to the model for each node restricting all variables associated with the connected edges. These constraints are formulated using quadratic terms of $x_i$ variables. As $x_i$ represents the colour of a node, $(1-2x_{i})(1-2x_{j})$ takes value $-1$ if and only if the two endpoints of edge $(i,j) \in E$ have different colours. Interpreting based on the concept of switching set, different values of $x_i$ variables associated with the two endpoints of edge $(i,j) \in E$ mean that the edge should be negated in the process of transforming to a negative minimal graph. The linearised formulation of the net-degree constraints using $x_i$ and $x_{ij}$ variables is provided in \eqref{eq10}. 

\begin{equation}\label{eq10}
\sum\limits_{j: (i,j)\in E \text{ or } (j,i)\in E} a_{ij} (1-2x_{i} -2x_{j} +4x_{ij}) \geq 0  \quad \forall i \in V
\end{equation}

Another structural property we observe is related to the edges making a cycle. According to the definition of the switching function \eqref{eq2}, switching one node negates all edges connected to that node. Because there are two edges connected to each node in a cycle, the negation states of edges making a cycle are not independent. To be more specific, the number of negated edges in each cycle of the graph must be even.

As listing all cycles of a graph is computationally intensive, this structural property can be applied to cycles of a limited length. For instance, we may apply this structural property to the edge variables making triangles in the graph. 
This structural property can be formulated as valid inequalities in Eq.\ \eqref{eq11} in which $T=\{(i,j,k)\in V^3 \mid (i,j),(i,k),(j,k) \in E \}$ contains ordered 3-tuples of nodes whose edges form a triangle. Note that $(x_{i} + x_{j} -2x_{ij})$ represents the negation state of the edges $(i,j) \in E$. The expression in Eq.\ \eqref{eq11} denotes the sum of negation states for the three edges $(i,j),(i,k),(j,k)$ making a triangle.

\begin{equation}\label{eq11}
\begin{split}
&x_{i} + x_{j} -2x_{ij}+x_{i} + x_{k} -2x_{ik}+x_{j} + x_{k} -2x_{jk}\\ 
&= 0 \text{ or } 2 \quad \forall (i,j,k) \in T
\end{split} 
\end{equation}

Eq.\ \eqref{eq11} can be linearised to Eq.\ \eqref{eq12} as follows. \textit{Triangle constraints} can be applied to the model as four constraints per triangle, restricting three edge variables and three node variables per triangle.

\begin{equation}\label{eq12}
\begin{split}
x_{i}+x_{jk} &\geq x_{ij} + x_{ik} 										  \quad \forall (i,j,k) \in T\\
x_{j}+x_{ik} &\geq x_{ij} + x_{jk}  											\quad \forall (i,j,k) \in T\\
x_{k}+x_{ij} &\geq x_{ik} + x_{jk}  											\quad \forall (i,j,k) \in T\\
1 + x_{ij} + x_{ik} + x_{jk} &\geq x_{i} + x_{j} + x_{k} \quad \forall (i,j,k) \in T
\end{split} 
\end{equation}



In order to speed up the model in \eqref{eq8}, we consider fixing a node colour to increase the root node objective function.
We conjecture the best node variable to fix is the one associated with the highest unsigned node degree. This constraint is formulated in \eqref{eq15} which our experiments show speeds up the branch and bound algorithm by increasing the lower bound.

\begin{equation}\label{eq15}
x_{k} = 1 \quad k= \text{arg} \max_{i \in V} d_i
\end{equation}

The complete formulation of the 0/1 linear model with further restrictions on the feasible space includes the objective function and core constraints in Eq.\ \eqref{eq8} and valid inequalities in Eq.\ \eqref{eq10}, Eq.\ \eqref{eq12}, and Eq.\ \eqref{eq15}. The model has $n+m$ binary variables, $m$ core constraints, and $n+4|T|+1$ additional constraints. 

Table \ref{tab1} provides a comparison of the three optimisation models based on their variables, constraints, and objective functions. In the next sections, we mainly focus on the 0/1 linear model solved in conjunction with the valid inequalities (additional constraints).

\begin{table}[h]
	\centering
	\caption{Comparison of the three optimisation models}
	\label{tab1}
	\begin{tabular}{p{2.5cm}p{2.5cm}p{2.5cm}l} 
		\hline\noalign{\smallskip}		& QCQP \eqref{eq4}  & UBQP \eqref{eq7}   & 0/1 linear model \eqref{eq8}       \\ 
		\noalign{\smallskip}\svhline\noalign{\smallskip}
		Variables       & $n$                & $n$       & $n+m$          \\
		Constraints     & $n$                & $0$       & $m$            \\
		Variable type   & continuous         & binary    & binary         \\
		Constraint type & quadratic       & -         & linear            \\
		Objective       & quadratic          & quadratic & linear         \\ \noalign{\smallskip}\hline\noalign{\smallskip}
	\end{tabular}
\end{table}

\section{Numerical results in random graphs} \label{s:results}
In this section, the frustration index of various random networks is computed by solving the 0/1 linear model \eqref{eq8} coupled with the additional constraints. We use Gurobi version 7 on a desktop computer with an Intel Corei5 4670 @ 3.40 GHz and 8.00 GB of RAM running 64-bit Microsoft Windows 7. The models were created using Gurobi's Python interface.

To verify our software implementation, we manually counted the number of frustrated edges given by our software's proposed node colouring for a number of test problems, and confirmed that this matched the frustration count reported by our software. These tests showed that our models and implementations were performing as expected.

\subsection{Performance of the 0/1 linear model on random graphs} \label{ss:perform}

In this subsection we discuss the time performance of the branch and bound algorithms for solving the 0/1 linear model. In order to evaluate the performance of the 0/1 linear model \eqref{eq8} coupled with the additional constraints, we generate 10 decent-sized Erd\H{o}s-R\'{e}nyi random graphs \cite{bollobas2001random} as test cases with various densities and percentages of negative edges. Results are provided in Table \ref{tab2} in which B\&B nodes stands for the number of branch and bound nodes (in the search tree of the branch and bound algorithm) explored by the solver.

\begin{table}[h]
	\centering
	\caption{Evaluation of the model in \eqref{eq8} based on random networks}
	\label{tab2}
	\begin{tabular}{p{1.5cm}p{1cm}p{1cm}p{1cm}p{1cm}p{1cm}p{1cm}p{1.7cm}l}
		\hline\noalign{\smallskip}
		TestCase	&	$n$	&	$m$	&	$m^-$	&	$\rho$	&	$\frac{m^-}{m}$	&	$Z^*$	&	B\&B nodes	&	time(s)\\
		\noalign{\smallskip}\svhline\noalign{\smallskip}
		1	&	65	&	570	&	395	&	0.27	&	0.69	&	189	&	5133	&	65.4\\
		2	&	68	&	500	&	410	&	0.22	&	0.82	&	162	&	4105	&	27.3\\
		3	&	80	&	550	&	330	&	0.17	&	0.60	&	170	&	11652	&	153.3\\
		4	&	50	&	520	&	385	&	0.42	&	0.74	&	185	&	901		&	22.4\\
		5	&	53	&	560	&	240	&	0.41	&	0.43	&	193	&	292		&	13.5\\
		6	&	50	&	510	&	335	&	0.42	&	0.66	&	178	&	573		&	13.8\\
		7	&	59	&	590	&	590	&	0.34	&	1.00	&	213	&	1831	&	46.0\\
		8	&	56	&	600	&	110	&	0.39	&	0.18	&	110	&	0		&	0.4\\
		9	&	71	&	500	&	190	&	0.20	&	0.38	&	155	&	6305	&	77.7\\
		10	&	80	&	550	&	450	&	0.17	&	0.82	&	173	&	12384	&	138.0\\
		\noalign{\smallskip}\hline\noalign{\smallskip}
	\end{tabular}
\end{table}

The results in Table \ref{tab2} show that random test cases based on Erd\H{o}s-R\'{e}nyi graphs with 500-600 edges can be solved to optimality in a reasonable time. The branching process for these test cases explores various numbers of nodes ranging between 0 and 12384. These numbers also depend on the heuristics that the solver uses automatically and solving the test cases for a second time often leads to a different number of branch and bound nodes.

\subsection{Impact of negative edges on the frustration index}

\begin{figure}[h]
	\centering
	\includegraphics[width=0.65\textwidth]{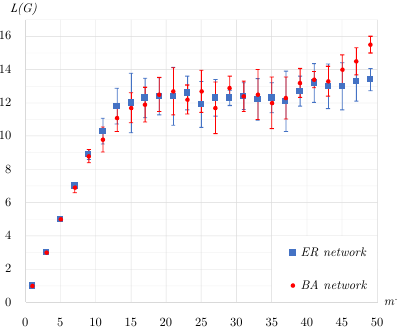}
	\caption{The frustration index in Erd\H{o}s-R\'{e}nyi (ER) networks with 15 nodes and 50 edges and Barab\'{a}si-Albert (BA) networks with 15 nodes and 50 edges and various number of negative edges (colour online)}
	\label{fig1}
\end{figure}

In this subsection we use both Erd\H{o}s-R\'{e}nyi and Barab\'{a}si-Albert random networks \cite{bollobas2001random} as synthetic data for calculation of the line index of balance. In this analysis, we use the same randomly generated graphs with different numbers of negative edges assigned by a uniform random distribution as test cases over 50 runs per experiment setting. Figure \ref{fig1} demonstrates the average and standard deviation of the line index of balance in these random signed networks with $n=15,m=50$. It is worth mentioning that we have observed similar results in other types of random graphs including small world, scale-free, and random regular graphs \cite{bollobas2001random}.

Figure \ref{fig1} shows similar increases in the line index of balance in the two graph classes as $m^-$ increases. It can be observed that the maximum frustration index is still smaller than $m/3$ for all graphs. This shows a gap between the values of the line index of balance in random graphs and the theoretical upper bound of $m/2$. It is important to know whether this gap is proportional to graph size and density.

\subsection{Impact of graph size and density on the frustration index}

\begin{figure*}[h]
	\centering
	\includegraphics[width=1\textwidth]{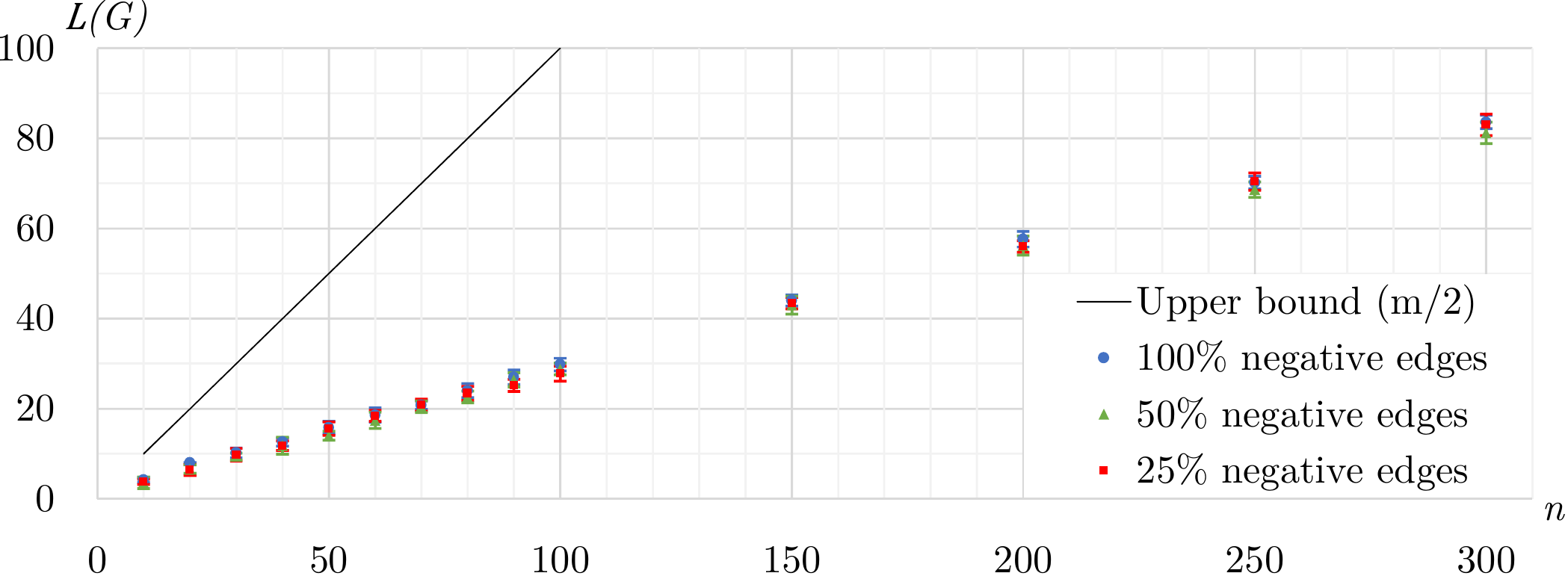}
	\caption{The frustration index in random 4-regular networks of different orders $n$ and decreasing densities (colour online)}
	\label{fig2}
\end{figure*}

In order to investigate the impact of graph size and density, 4-regular random graphs with a constant fraction of randomly assigned negative edges are analysed averaging over 50 runs per experiment setting. The frustration index is computed for 4-regular random graphs with 25\%, 50\%, and 100\% negative edges and compared with the upper bound $m/2$. Figure \ref{fig2} demonstrates the average and standard deviation of the frustration index where the degree of all nodes remains constant, but the density of the 4-regular graphs, $\rho=4/n-1$, decreases as $n$ and $m$ increase.

An observation to derive from Figure \ref{fig2} is the similar frustration index values obtained for networks of the same sizes, even if they have different percentages of negative edges. It can be concluded that starting with an all-positive graph (which has a frustration index of $0$), making the first quarter of graph edges negative increases the frustration index much more than making further edges negative. Future research is required to get a better understanding of how the frustration index and minimum deletion sets change when the number of negative edges is increased (on a fixed underlying structure). Another observation is that the gap between the frustration index values and the theoretical upper bound increases with increasing $n$. 

\section{Numerical results in real signed networks} \label{s:real}
In this section, the frustration index is computed in nine real networks by solving the 0/1 linear model \eqref{eq8} using Gurobi version 7 on a desktop computer with an Intel Corei5 4670 @ 3.40 GHz and 8.00 GB of RAM running 64-bit Microsoft Windows 7.

There are well studied signed social network datasets representing communities with positive and negative interactions and preferences. Read's dataset for New Guinean highland tribes \cite{read_cultures_1954} and Sampson's dataset for monastery interactions \cite{sampson_novitiate_1968} which we denote respectively by G1 and G2. We also use graphs inferred from datasets of students' choice and rejection, denoted by G3 and G4 \cite{newcomb_acquaintance_1961,lemann_group_1952}. A further explanation on the details of inferring signed graphs from choice and rejection data can be found in \cite{aref2015measuring}. Moreover, a larger signed network, denoted by G5, is inferred by \cite{neal_backbone_2014} through implementing a stochastic degree sequence model on Fowler's data on Senate bill co-sponsorship \cite{fowler_legislative_2006}.

As well as the signed social network datasets, large scale biological networks can be analysed as signed graphs. There are four signed biological networks analysed by \cite{dasgupta_algorithmic_2007} and \cite{iacono_determining_2010}. Graph G6 represents the gene regulatory network of \textit{Saccharomyces cerevisiae} \cite{Costanzo2001yeast} and graph G7 is related to the gene regulatory network of \textit{Escherichia coli} \cite{salgado2006ecoli}. The Epidermal growth factor receptor pathway \cite{oda2005} is represented as graph G8. Graph G9 represents the molecular interaction map of a macrophage \cite{oda2004molecular}. For more details on the four biological datasets, one may refer to \cite{iacono_determining_2010}. The data for real networks used in this study is publicly available on the \href{https://figshare.com/articles/Signed_networks_from_sociology_and_political_science_biology_international_relations_finance_and_computational_chemistry/5700832}{Figshare} research data sharing website \cite{Aref2017data}.

We use $G_r=(V,E,\sigma_r)$ to denote a reshuffled graph in which the sign function $\sigma_r$ is a random mapping of $E$ to $\{-1,+1\}^m$ that preserves the number of negative edges. The reshuffling process preserves the underlying graph structure. The numerical results on the frustration index of our nine signed graphs and reshuffled versions of these graphs are shown in Table \ref{tab3} where, for each graph $G$, the average and standard deviation of the line index of balance in 500 reshuffled graphs, denoted by $L(G_r)$ and $\text{SD}$, are also provided for comparison. 

\begin{table}[h]
	\centering
	\caption{The frustration index in various signed networks}
	\label{tab3}
	\begin{tabular}{p{1.5cm}p{1cm}p{1cm}p{1cm}p{1cm}p{2.5cm}l}
		\hline\noalign{\smallskip}
		Graph & $n$ & $m$ & $m^-$ & $L(G)$ & $L(G_r) \pm \text{SD}$ & Z score \\ 
		\noalign{\smallskip}\svhline\noalign{\smallskip}
		G1    & 16  & 58  & 29    & 7    & $14.65 \pm 1.38$ &  -5.54 \\
		G2    & 18  & 49  & 12    & 5    & $9.71  \pm 1.17$ &  -4.03 \\
		G3    & 17  & 40  & 17    & 4    & $ 7.53 \pm 1.24$ &  -2.85 \\
		G4    & 17  & 36  & 16    & 6    & $ 6.48 \pm 1.08$ &  -0.45 \\
		G5    & 100 & 2461& 1047  & 331  & $ 965.6\pm 9.08$& -69.89 \\ 
		G6    & 690 & 1080& 220   &  41  & $ 124.3\pm 4.97$& -16.75 \\ 
		G7    & 1461& 3215& 1336  & 371  & $ 653.4\pm 7.71$& -36.64 \\ 	
		G8    & 329 & 779 & 264   & 193  & $ 148.96\pm 5.33$&   8.26 \\ 
		G9    & 678 & 1425&  478  & 332  & $ 255.65\pm 8.51$&  8.98 \\ 	
		\noalign{\smallskip}\hline\noalign{\smallskip}	
	\end{tabular}
\end{table}

Although the signed networks G1 -- G7 are not balanced, the relatively small values of $L(G)$ suggest a low level of frustration in some of the networks. G1 -- G7 exhibit a level of frustration lower than what is expected by chance (obtained through random allocation of signs to the unsigned graph), while the opposite is observed for G8 and G9.

\begin{figure}[h]
	\centering
	\begin{subfigure}[t]{0.45\textwidth}
		\centering
		\includegraphics[height=1.5in]{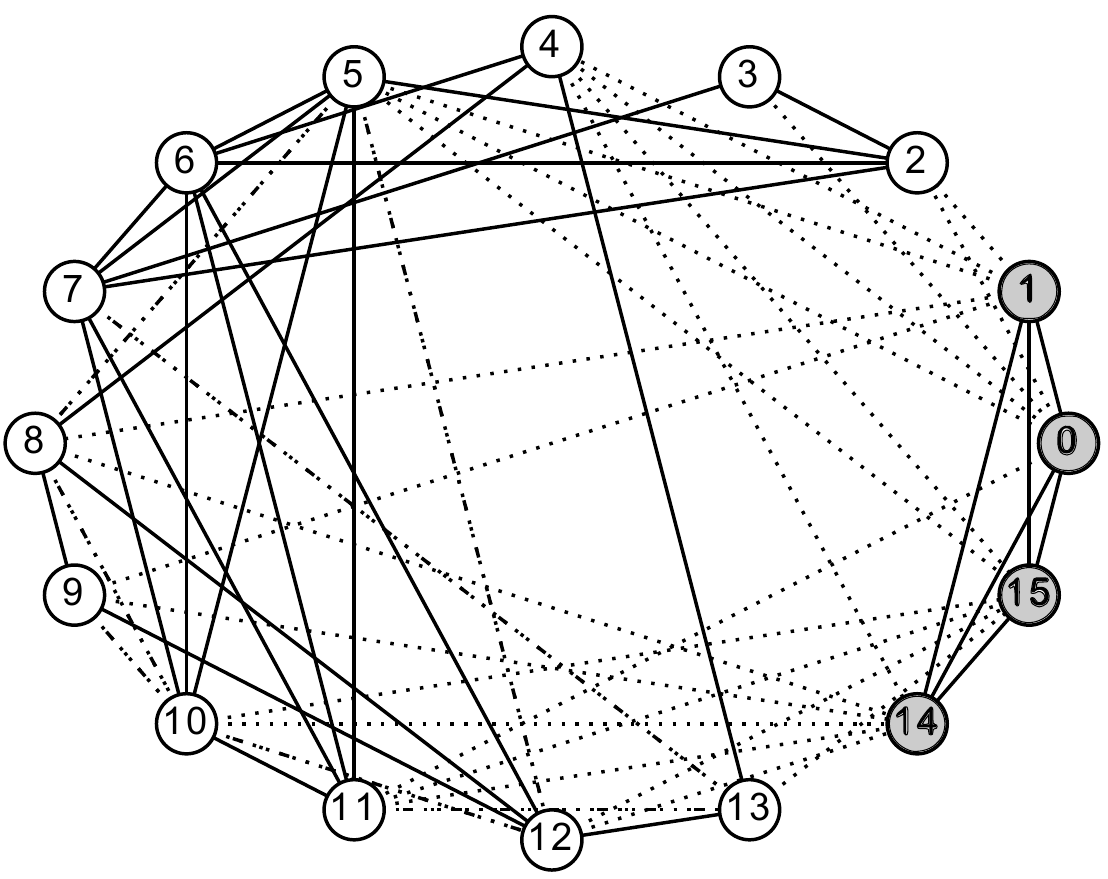}
		\caption{Highland tribes network (G1), a signed network of 16 tribes of the Eastern Central Highlands of New Guinea \cite{read_cultures_1954}. Minimum deletion set comprises 7 negative edges.}
	\end{subfigure}%
	~ \quad
	\begin{subfigure}[t]{0.45\textwidth}
		\centering
		\includegraphics[height=1.5in]{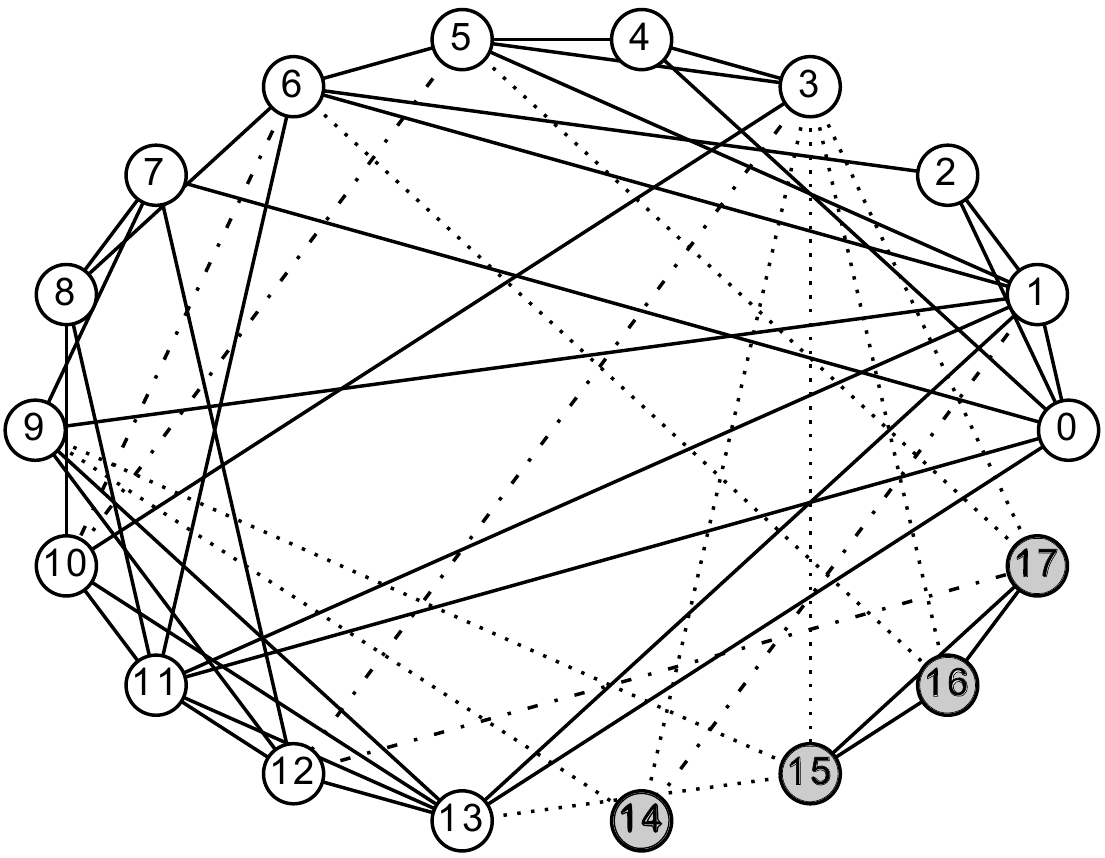}
		\caption{Monastery interactions network (G2) of 18 New England novitiates inferred from the integration of all positive and negative relationships \cite{sampson_novitiate_1968}. Minimum deletion set comprises 2 positive and 3 negative edges.}
	\end{subfigure}
	~ 
	\begin{subfigure}[t]{0.45\textwidth}
		\centering
		\includegraphics[height=1.5in]{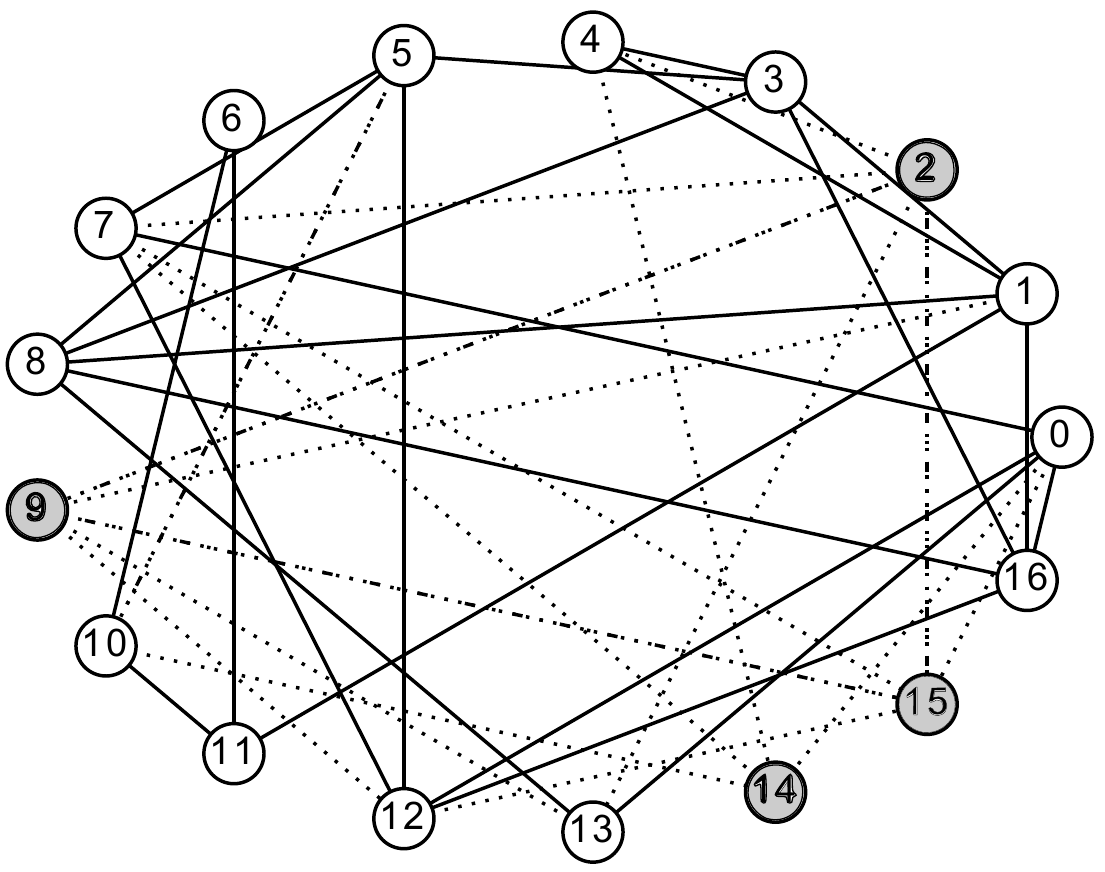}
		\caption{Fraternity preferences network (G3) of 17 boys living in a pseudo-dormitory inferred from ranking data of the last week in \cite{newcomb_acquaintance_1961}. Minimum deletion set comprises 4 negative edges.}
	\end{subfigure}
	~ \quad
	\begin{subfigure}[t]{0.45\textwidth}
		\centering
		\includegraphics[height=1.5in]{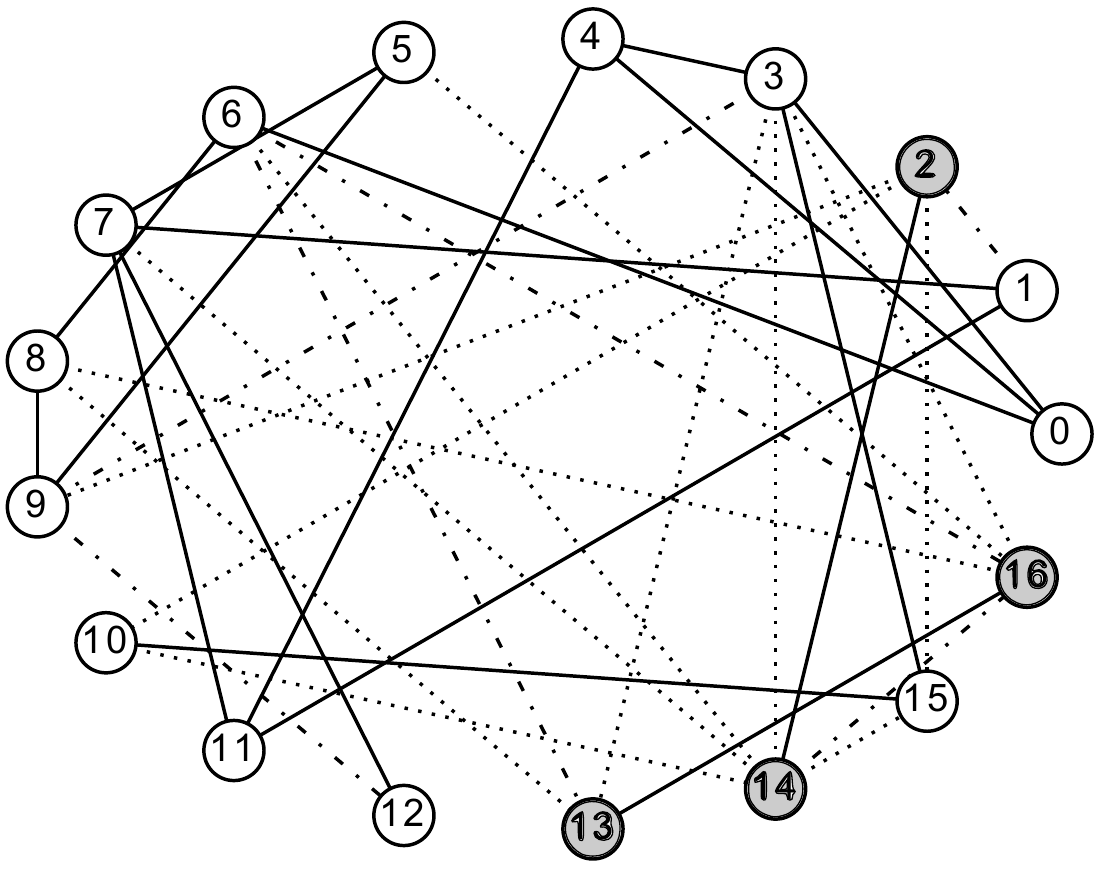}
		\caption{College preferences network (G4) of 17 girls at an Eastern college inferred from ranking data of house B in \cite{lemann_group_1952}. Minimum deletion set comprises 3 positive and 3 negative edges.}
	\end{subfigure}
	\caption{The frustrated edges represented by dotdash lines for four small signed networks inferred from the sociology datasets}
	\label{fig3}
\end{figure}

Figure \ref{fig3} shows how the small signed networks G1 -- G4 can be made balanced by negating (or removing) the edges on a minimum deletion set. Dotted lines represent negative edges, solid lines represent positive edges, and frustrated edges are indicated by dotdash lines regardless of their original signs. The node colourings leading to the minimum frustration counts are also shown in Figure \ref{fig3}. Note that it is pure coincidence that there are an equal number of nodes coloured black for each graph G1 -- G4 in Figure \ref{fig3}. Visualisations of graphs G1 -- G4 without node colours and minimum deletion sets can be found in \cite{aref2015measuring}.

In order to be more precise in evaluating the relative levels of frustration in G1 -- G9, we have implemented a very basic statistical analysis using Z scores, where $Z={(L(G)-L(G_r))}/{\text{SD}}$. The Z scores, provided in the right column of Table \ref{tab3}, show how far the frustration index is from the values obtained through random allocation of signs to the fixed underlying structure (unsigned graph). 
Negative values of the Z score can be interpreted as a lower level of frustration than the value resulting from a random allocation of signs. This comparison allows us to say that the level of frustration is very low for G5, G6, and G7, low for G1 and G2, and very high for G8 and G9.

Various performance measures for the 0/1 linear model \eqref{eq8} coupled with the additional constraints for solving G1 -- G9 are provided in Table \ref{tab4}. 

\begin{table}[h]
	\centering
	\caption{Performance measures for the model in \eqref{eq8} based on real networks}
	\label{tab4}
	\begin{tabular}{p{1.5cm}p{1.5cm}p{3cm}p{2cm}l}
		\hline\noalign{\smallskip}
		Graph & $L(G)$ & Root node objective & B\&B nodes 		& Solve time (s) \\ 
		\noalign{\smallskip}\svhline\noalign{\smallskip}
		G1    & 7      &    4.5              &     0                &   0.03         \\
		G2    & 5      &    0              &      0               &   0.04         \\
		G3    & 4      &    2.5              &      0               &   0.02         \\
		G4    & 6      &    2              &     0                &  0.04          \\
		G5    & 331    &   36.5               &    0                 &  78.67          \\
		G6    & 41     &     3             &       0              &   0.28         \\
		G7    & 371    &     21.5             &    1085                 &    27.22        \\
		G8    & 193    &     17             &   457                  &    0.72        \\
		G9    & 332    &       14.5           &      1061               &    1.92        \\ 
		\noalign{\smallskip}\hline\noalign{\smallskip}
	\end{tabular}
\end{table}

We compare the quality and solve time of our exact algorithm with that of recent heuristics and approximations implemented on the datasets.
Table \ref{tab5} provides a comparison of the 0/1 linear model \eqref{eq8} with other methods in the literature.

\begin{table}[h]
	\centering
	\caption{Comparison of the solution and solve time against models in the literature}
	\label{tab5}
\begin{tabular}{lp{1.5cm}p{3cm}p{3cm}l}
	\hline\noalign{\smallskip}
	&\multicolumn{1}{l}{Graph}              & \multicolumn{1}{l}{H\"{u}ffner et al.\ \cite{huffner_separator-based_2010}} & \multicolumn{1}{l}{Iacono et al.\ \cite{iacono_determining_2010}} & \multicolumn{1}{l}{0/1 linear model} \\ 
	\noalign{\smallskip}\svhline\noalign{\smallskip}
	\multirow{4}{*}{Solution} 

	& G6 &  41          			                & 41                                & 41                           \\
	& G7 & Not converged                            & {[}365, 371{]}                    & 371                      \\
	& G8 &  210				                        & {[}186, 193{]}                    & 193                        \\
	& G9 &  374         			                & {[}302, 332{]}                    & 332                        \\ \cline{2-5} 
	\multirow{4}{*}{Time}      
	& G6 & 60 s	                                    & A few minutes                     & 0.28 s                     \\
	& G7 & Not converged                            & A few minutes                     & 27.22 s                   \\
	& G8 & 6480 s                                   & A few minutes                     & 0.72 s                      \\
	& G9 & 60 s                                     & A few minutes                     & 1.92 s                     \\ 
	\noalign{\smallskip}\hline\noalign{\smallskip}
\end{tabular}
\end{table}

H\"{u}ffner et al.\ have previously investigated frustration in G6 -- G9 suggesting a data reduction scheme and (an attempt at) an exact algorithm \cite{huffner_separator-based_2010}. Their suggested data reduction algorithm can take more than 5 hours for G6, more than 15 hours for G8, and more than 1 day for G9 if the parameters are not perfectly tuned \cite{huffner_separator-based_2010}. Their algorithm coupled with their data reduction scheme and heuristic speed-ups does not converge for G7 \cite{huffner_separator-based_2010}. In addition to these solve time and convergence issues, their algorithm provides $L(G8)=210, L(G9)=374$, both of which are incorrect based on our results. 

Iacono et al.\ have investigated frustration in G6 -- G9 \cite{iacono_determining_2010}. Their heuristic algorithm provides upper and lower bounds for G6 -- G9 with a 100\%, 98.38\%, 96.37\% and 90.96\% ratio of lower to upper bound respectively. Regarding solve time, they have only mentioned that their heuristic requires a fairly limited amount of time (a few minutes on an ordinary PC).

While data reduction schemes \cite{huffner_separator-based_2010} take up to 1 day for these datasets and heuristic algorithms \cite{iacono_determining_2010} only provide bounds with up to 9\% gap from optimality, our 0/1 linear model solves each of the 9 datasets to optimality in less than a minute.

\section{Conclusion} \label{s:conclu}

This study focuses on frustration index as a measure of balance in signed networks and the findings may well have a bearing on the applications of the line index of balance in the other disciplines \cite{aref2017balance}. The present study has suggested a novel method for computing a measure of structural balance that can be used for analysing dynamics of signed networks. It contributes additional evidence that suggests signed social networks and biological gene regulatory networks exhibit a relatively low level of frustration (compared to the expectation when allocating signs at random). On similar lines of research, we have undertaken a follow-up study with more focus on operations research aspects of this topic \cite{aref2016exact}.

This study has a number of important implications for future investigation. The optimisation model introduced can make network dynamics models more consistent with the theory of structural balance \cite{antal_dynamics_2005}. 
To be more specific, many sign change simulation models that allow one change at a time use the number of balanced triads in the network as a criterion for transitioning towards balance. These models may result in stable states that are not balanced, like jammed states and glassy states \cite{marvel_energy_2009}. This contradicts not only the instability of unbalanced states, but the fundamental assumption that networks gradually move towards balance. Deploying decrease in the frustration index as the criterion, the above-mentioned states might be avoided resulting in a more realistic simulation of signed network dynamics that is consistent with structural balance theory and its assumptions.

\begin{acknowledgement}
The authors are grateful for the extremely valuable comments of the anonymous reviewers that have prevented incorrect attributions in the literature review section and helped improve the discussions in this chapter.
\end{acknowledgement}

\bibliographystyle{spmpsci}
\bibliography{refs}

\end{document}